\documentclass[aps,twocolumn,prl,superscriptaddress,floatfix,showpacs]{revtex4}
\usepackage{graphicx}
\usepackage{color}
\usepackage{epstopdf}
\newcommand{\bi}{\bibitem}
\newcommand{\ba}{\begin{eqnarray}}
\newcommand{\ea}{\end{eqnarray}}
\def\ct{\cite}

\begin{document}

\title{Role of marginality in quantum fidelity and Loschmidt echo:  Dirac points in 2-D}
\author{Aavishkar A. Patel}
\author{Shraddha Sharma}
\author{Amit Dutta}
\affiliation{Department of Physics, Indian Institute of Technology Kanpur, Kanpur 208016, India}

\begin{abstract}
We investigate the effect of marginality on the  ground state fidelity and Loschmidt echo. For this purpose, we  study 
the above quantities near the  quantum critical point (QCP) of the two-dimensional (2-D) Dirac Hamiltonian
in the presence of a mass term which is tuned to zero at the Dirac point. An ideal example would be that of the low-energy carriers in graphene in which a mass term opens up a band gap.  This happens to be a marginal situation where the behavior of the fidelity and the echo is markedly different as compared to that in the one-dimensional case. We encounter this marginal behavior near the Dirac point, which is displayed in the absence of a sharp dip in the ground state fidelity (or equivalently in  the logarithmic scaling of the fidelity susceptibility). Most importantly, there is also a logarithmic correction to the proposed scaling of the fidelity in the thermodynamic limit which can not
be a priori anticipated from the predicted scaling form. Interestingly, a sharp dip in the ground state Loschmidt echo is also found to be absent near this QCP, which is again a consequence of  the marginality. We also explain the absence of a sharp dip in both  the fidelity and the Loschmidt echo close to the QCP in dimensions greater than two. \end{abstract}

\pacs{64.70.Tg, 03.65.Pm}

\maketitle

\section{I. Introduction}

The scaling behavior of different thermodynamic variables close to a critical point is strongly influenced by a marginal situation. In this letter, we study how marginality influences the ground state fidelity and the ground state Loschmidt echo close to a quantum
critical point (QCP) \ct{sachdev99,chakrabarti96,sondhi97}. We study the above quantities for two-dimensional (2-D) Dirac Hamiltonians (DHs) close to the
QCP (which happens to be a Dirac point (DP) with a linear dispersion) for which exact analytical calculations (in all limits) are possible.

It is worth mentioning that the one-dimensional (1-D) and 2-D DHs have found a wide range of applications in quantum condensed matter systems in recent years~\cite{castroneto09,hasan10,qi11}. The low-energy physics of graphene~\ct{novoselov2005two}, and of the bulk states in 2-D topological insulators is described by a 2-D DH~\cite{kane05,bernevig06}, and the 1-D edge states existing in a 2-D topological insulator system are described by an effective 1-D DH~\cite{qi11}. QCPs of graphene (the gapless to gapped transition induced by a mass term) are 2-D  DPs with linear dispersions. Experimentally, a gap can be opened in the otherwise gapless linear band structure of graphene through several methods e.g., by the application of an external electric field to the graphene sheet~\cite{quhe2012tunable}; this is mimicked by adding a mass term to the DH which vanishes at the gapless QCP. The experimental prospect of  tuning of parameters controlling these quantum phase transitions (QPTs) in optical~\cite{lee09} and photonic lattices~\cite{ochiai09} have made the study of these Hamiltonains also from the viewpoint of quantum information timely and necessary.

In this letter, we study the ground state quantum  fidelity \ct{zanardi06,zhou08fidelity,zhou08ground,sacramento11} of a 2-D DH and show the difference in results as compared to the 1-D case. We encounter marginality in the behavior of the ground state fidelity near the DP which shows up in a prominent way both in the scaling of the fidelity susceptibility and the scaling of the fidelity in the thermodynamic limit. Although the marginal situation is indicated in the scaling of the fidelity susceptibility, it is not a priori obvious that it will influence the
scaling in the thermodynamic limit as well (We note that the fidelity susceptibility is proportional to the density of defects generated by a sudden quench across a QCP~\ct{grandi10,patel12}). We also study the scaling behavior of the ground state Loschmidt echo (LE)\ct{quan06} close to the QCP of the 2-D DH. It shows a pronounced drop near the QCP in the 1-D case but not in the 2-D case.  This occurs again due to the marginality of the 2D DH. We also discuss  the behavior of the fidelity and the LE for a $d$-dimensional DP and explain how the marginal case is different from the other situations.  In short, we shall show below that in the marginal situation the behavior of the fidelity and the LE is strikingly different and is dependent on an upper cut-off in momentum space. To the best of our knowledge, this marginal situation is not explored in details; the 2-D DH which  is  an integrable model as well as reducible to a $2 \times 2$ form turns out to
be an ideal candidate to explore the role of marginality. Moreover, it enables us to study the fidelity and the echo close to
a topological QPT.

The letter is organized in the following fashion. In Sec. II, we introduce a tight binding Hamiltonian  on a honeycomb lattice and arrive at a massive 2-D DH. Sec. III is devoted to the studies of the fidelity susceptibility and the fidelity in the thermodynamic limit for the 2-D DH while in Sec. IV we present the studies of the ground state LE.  The results obtained in Secs. III and IV are generalized  in Sec. V for a DP in $d$ dimensions. We summarize the main results of the paper and their implications in the concluding section.

\section{II. Model}
The 2-D DH with a mass term can be realised in condensed matter systems as the effective low-energy Hamiltonian of particles hopping on a honeycomb lattice with unequal sublattice potentials:
\begin{equation}
H=-t\sum_{\vec{n},\vec{\delta}}\left(a^{\dag}_{\vec{n}}b_{\vec{n}+\vec{\delta}}+b^{\dag}_{\vec{n}+\vec{\delta}}a_{\vec{n}}\right)+\sum_{\vec{n}}\left(\mu_{1}a^{\dag}_{\vec{n}}a_{\vec{n}}+\mu_{2}b^{\dag}_{\vec{n}}b_{\vec{n}}\right),
\end{equation}
where $\vec{n}$ labels the bonds of the honeycomb lattice connecting sites on sublattices $a$ and $b$ and $\vec \delta$'s are the vectors connecting the  bonds involving the
nearest neighbor lattice sites; the case $\mu_1 =\mu_2$ represents the gapless graphene Hamiltonian \ct{castroneto09}. The low energy spectrum of this Hamiltonian occurs in two distinct valleys at the corners of a hexagonal Brillouin zone, and is described by a $4\times4$ block diagonal Hamiltonian with two $2\times2$ blocks~\cite{castroneto09}:
\begin{equation}
H_{eff}= \left(
\begin{array}{clrr} 
    \mathcal H_D (\vec{k}) &0 \\ 
    0  &\mathcal H_D^*(-\vec{k})
\end{array}
\right);
\end{equation}
after appropriately rescaling units and ignoring contributions proportional to the identity matrix, one arrives at
\begin{equation}
\mathcal H_D = \left(
\begin{array}{clrr} 
    m &k_x-ik_y \\ 
    k_x+ik_y  &- m
\end{array}
\right).
\label{Eq:Ham_Massive_Dirac}
\end{equation}
The momentum $\vec{k}$ is measured with respect to the corners of the Brillouin zone. The particles in the two distinct valleys are thus independently described by massive 2-D DHs, where the parameter $m$ is the Dirac mass. This produces two bands with dispersions $E_{\vec{k}}^{\pm}=\pm\sqrt{k_x^2+k_y^2+m^2}$, and displays a QCP (DP) at $m=0$ (with the associated correlation length exponent $\nu=1$) where the gap between the two bands vanishes (such as in graphene). This is a topological phase transition characterized by a change in the Chern numbers (or equivalently, the Berry phase of modes near the critical mode $\vec k=0$) of the bands as $m$ goes from negative to positive values via the QCP.

\section{III. ground state fidelity}
The ground state quantum fidelity~\cite{zanardi06} ($F$), which measures the overlap between many-body ground states at slightly different values of a parameter $m$ of the Hamiltonian, is an important tool for detecting quantum phase transitions~\cite{,zanardi06,zhou08fidelity,gu10,venuti07,dutta10,zhou08ground,damski12}: one can use the expansion
$F=|\langle \psi(m+\delta)| \psi(m) \rangle|^2 = 1-(\delta^2/2) L^d \chi_F (m) +  ...$, where $\delta (\to 0)$ is a small change  in the parameter $m$ and $L$ is the linear dimension of the $d$-dimensional system.  One also assumes a small system size limit and $\chi_F(m)$ is defined to be the fidelity susceptibility density at $m$. The quantum fidelity generally displays a marked drop as $m$ approaches the value $m_c$ at which the Hamiltonian has a QCP. Likewise, $\chi_F$ usually scales in a universal power-law fashion away from the QCP~\ct{venuti07,gu10,dutta10,gritsev09,schwandt09,mukherjee11}  as $|m|^{\nu d-2}$;  near the QCP, $\chi_F \sim L^{2/\nu-d}$. We explore the marginal case with $\nu d =2$, using the example of the 2-D  DH with a mass term (since $\nu=1$), while the 1-D case has been found to satisfy the proposed scaling relation  \ct{mukherjee11a}.  It has been shown for a one-dimensional quantum spin model  with $\nu d >2$ that neither the fidelity nor the fidelity susceptibility  is effective in detecting the critical point \ct{thakurati11}. 

On the other hand, in the thermodynamic limit ($L\rightarrow\infty$ and small but finite $\delta$), one cannot use the fidelity susceptibility approach~\ct{rams11}.
It has been conjectured that in this limit,  $\ln F  \sim - \delta^{\nu d} L^d$ at the QCP; this scaling  has been verified for the 1-D DH \ct{mukherjee11a}. We shall show below that the marginality shows up in the scaling of both  the fidelity susceptibility and the fidelity in the thermodynamic limit.

One can get an exact expression for the ground state fidelity $F$ using the two-level nature of the Hamiltonian (\ref {Eq:Ham_Massive_Dirac}). The ground state with a given number of modes is
\begin{equation}
|\psi(m)\rangle=\bigotimes_{2\pi/L<|\vec{k}|<k_{max}} |\psi_{k}^{-}(m)\rangle,
\label{Eq:groundstate}
\end{equation}
where
\begin{equation} 
|\psi_{\vec{k}}^{\pm}(m)\rangle = \frac{k}{\sqrt{2k^2+2m^2\pm2m\sqrt{k^2+m^2}}}
\left(
\begin{array}{clrr} 
    \frac{m \pm \sqrt{k^2+m^2}}{k_x+i k_y} \\ 
    								1
\end{array}
\right).
\label{Eq:spinor}
\end{equation} 
The fidelity is thus given by a product over all relevant modes up to a certain upper cutoff $k_{max}\leq2\pi/a_0$ (where $a_0$ is the lower length scale in the problem, usually provided by the lattice spacing), which plays an important role that is subsequently illustrated. The effect of finite system size $L$ is implemented in this continuum model by using periodic boundary conditions, providing a lower cutoff on the $\vec{k}$ modes given by $2\pi/L$. We find

\begin{eqnarray}
F=\prod_{2\pi/L\leq |\vec{k}|\leq k_{max}} |\langle \psi_{\vec{k}}(m+\delta)| \psi_{\vec{k}}(m) \rangle|^2, \nonumber \\
\ln F= \left(\frac{L}{2\pi}\right)^2\int_{2\pi/L}^{k_{max}} d^2\vec{k}~\ln |\langle \psi_{\vec{k}}(m+\delta)| \psi_{\vec{k}}(m) \rangle|^2.
\label{Eq:Fidelity}
\end{eqnarray}

\begin{figure}[h]
\includegraphics[width=8.6cm]{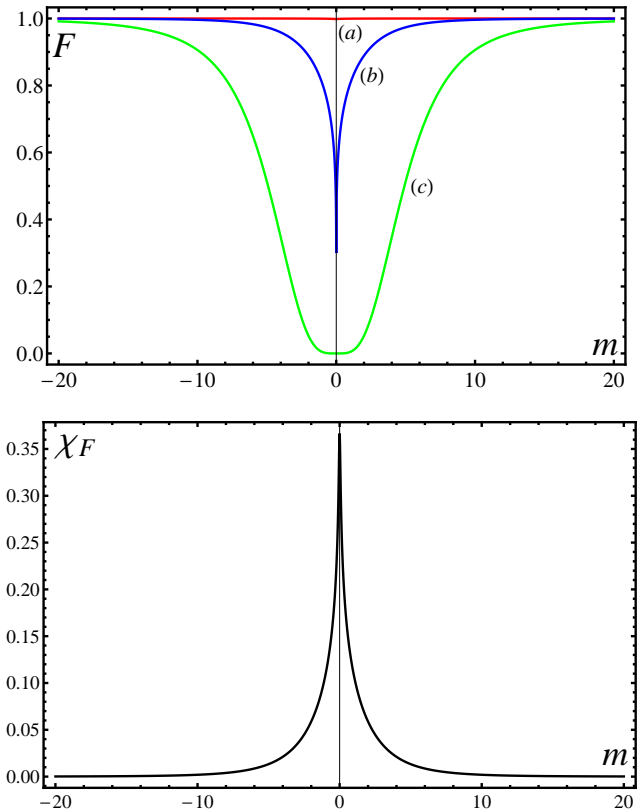}
\caption{(Top panel) The ground state fidelity calculated for  $\delta=0.001$ and $k_{max}=2\pi$, with (a) $L=100$ (fidelity susceptibility limit), (b) $L=2,000$ (an intermediate case), (c) $L=10,000$ (thermodynamic limit). In all these cases, there is no sharp dip at the QCP, and the fidelity starts to drop from unity when $|m|\sim k_{max}$. (Bottom panel) The fidelity susceptibility density, as obtained from (\ref{Eq:Fidelity_Massive_Dirac}), for the case (a) above, in which it is meaningful. This shows a peak at the QCP ($m=0$).}
\label{Fig:Fidelity_Massive_Dirac}
\end{figure}

Let us first consider the limit of  $\delta\ll1/L$, with finite $L$ and $k_{max}$, in which the notion of the fidelity susceptibility is meaningful; we find
\begin{eqnarray}
F&=&1-\frac{\delta^2L^2}{16\pi}(\frac{m^2}{k_{max}^2+m^2} - \nonumber \\ 
&~&\frac{L^2m^2}{4\pi^2+L^2m^2}+\ln\frac{L^2(k_{max}^2+m^2)}{L^2m^2+4\pi^2})+...
\label{Eq:Fidelity_Massive_Dirac}
\end{eqnarray}
The fidelity susceptibility density  then satisfies the following scaling relations: 
\begin{eqnarray}
\chi_{F} &\sim& \frac{1}{4\pi}\ln\left(\frac{L k_{max}}{4\pi} \right),  ~ m \ll \frac 1{L} \ll k_{max} \nonumber \\
\chi_{F}&\sim& \frac{1}{4\pi} \ln\left(\frac{k_{max}}{m}\right), ~ k_{max} \gg m \gg \frac 1{L},
\end{eqnarray}
while the first relation is valid close to the QCP, the second one corresponds to the situation away from it.
These logarithmic scaling forms clearly indicate the marginality of the situation and the DH allows us to derive
the scaling form exactly. In the fidelity susceptibility  limit, there is a nominal drop in the fidelity close to the QCP (see Fig.~\ref{Fig:Fidelity_Massive_Dirac}). 

We now address the role of marginality in the scaling of the fidelity in the thermodynamic limit, which
is not immediately obvious from the relation given above. 
Here, the thermodynamic limit refers to the situation in which $k_{max}\gg\delta\gg1/L$. Using the exact expression of
 the fidelity obtained through (\ref{Eq:Fidelity}), we find, at the QCP,
\begin{equation}
\ln F = \frac{L^2\delta^2}{8\pi} \ln \left(\frac{2\delta}{k_{max}}\right).
\end{equation}
We therefore have a marginal logarithmic correction to the expected $\delta^2$ scaling of the fidelity even in the thermodynamic limit. Away from the QCP, we have
\begin{equation}
\ln F = \frac{L^2\delta^2}{16\pi} \left[\frac{k_{max}^2}{k_{max}^2+m^2}+\ln\left(\frac{m^2}{k_{max}^2+m^2}\right)\right].
\label{eq_fid_thermo_away}
\end{equation}
Comparing Eq.~(\ref{eq_fid_thermo_away}) with Eq.~(\ref{Eq:Fidelity_Massive_Dirac}), we find that $\ln F$ scales in a
similar fashion as $\chi_F$ (though the latter is  defined only in the limit of $\delta \ll 1/L$) as predicted \ct{rams11}.  As $k_{max} \gg \delta$ in all limits, the crossover from the fidelity susceptibility limit to  the thermodynamic limit occurs when $\delta$ becomes of the order $1/L$. This crossover from the  susceptibility limit to the thermodynamic limit as the linear dimension $L$  as well as  $k_{max}$ are increased is shown in Fig.~2.

\begin{figure}[h]
\includegraphics[width=8.6cm]{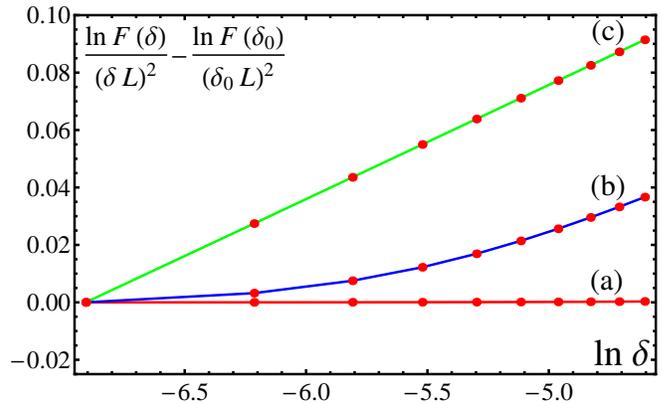}
\caption{Crossover of the behavior of the ground state fidelity at the QCP ($m=0$) from the (a) non-thermodynamic (fidelity susceptibilty) limit, with $L=100$ and $k_{max}=0.2\pi$ to the thermodynamic limit (c), with $L=100,000$ and $k_{max}=200\pi$, via an intermediate case (b) with $L=2,000$ and $k_{max}=4\pi$. Here $\delta$ ranges from $\delta=\delta_
{0}=0.001$ to $\delta=0.01$. It is clear from the figure that $\ln F$ scales as $\delta^2$ in the susceptibility limit and $\delta^2\ln \delta$ in the thermodynamic limit.}
\label{Fig:Fidelity_Crossover}
\end{figure}

\section {IV. Ground State LE}
The LE~\cite{quan06} $\mathcal{L}(t)$ for a many-body ground state $|\psi(m)\rangle$, evolved with two different Hamiltonians, one at $m$ and the other at $m+\delta$ is the measure of the overlap given by
\begin{equation}
\mathcal{L}(t) = |\langle\psi(m)|e^{iH(m)t}e^{-iH(m+\delta)t}|\psi(m)\rangle|^2.
\label{Eq:Echo}
\end{equation}
Usually, the LE, which can be shown to be related to the static fidelity \ct{zanardi06} shows a sharp dip in the vicinity of a QCP~\ct{echo,damski11}; this drop  has also been detected experimentally using NMR quantum simulator \ct{zhang09}.

\begin{figure}[h]
\includegraphics[width=8.6cm]{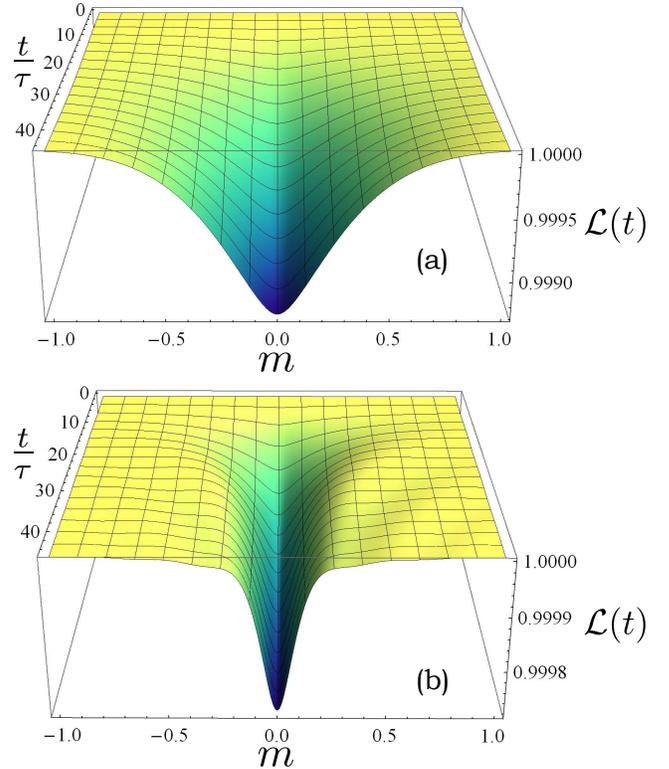}
\caption{(a) The ground state LE of the massive 2-D DH. Like the fidelity, the LE also does not show a sharp dip near the QCP ($m=0$). It has a Gaussian early time decay for values of $m$ close to the QCP. Here $\delta=0.001$, $L=100$ and $k_{max}=0.2\pi$. The scale $\tau$ is set by the early time decay $\mathcal{L}(t)=e^{-\delta^2(t/\tau)^2}$. (b) The same for the massive 1-D DH, with the same values of $\delta$, $L$ and $k_{max}$. The dip near the QCP is  sharp.}
\label{Fig:Echo_Massive_Dirac}
\end{figure}

We calculate the LE for the massive 2-D  DH  for arbitrary $m$ and $\delta$.  The evolution with the Hamiltonian $H(m+\delta)$ for a mode $|\psi_{\vec{k}}^{-}(m)\rangle$ of the ground state at  $m$ (i.e., of the Hamiltonian $H(m))$ is given by
\begin{eqnarray}
&&e^{-i H(m+\delta) t} |\psi_{\vec{k}}^{-}(m)\rangle = \sum_{\pm} \langle \psi_{\vec{k}}^{\pm}(m+\delta)|\psi_{\vec{k}}^{-}(m)\rangle  \nonumber \\
&&\times~e^{-i E_{\vec{k}}^{\pm}(m+\delta) t} |\psi_{\vec{k}}^{\pm}(m+\delta)\rangle.
\end{eqnarray}
Using (\ref{Eq:groundstate}), (\ref{Eq:spinor}) and (\ref{Eq:Echo}), we obtain
\begin{equation}
\mathcal{L}(t) = \prod_{2\pi/L<|\vec{k}|<k_{max}}\left[1-\frac{k^2 \delta^2 \sin^2(\sqrt{k^2+m^2}t) }{(k^2+m^2)(k^2+(m+\delta)^2)}\right].
\label{Eq:Echo_Massive_Dirac}
\end{equation}
This can be evaluated in a similar manner as done for the fidelity, by taking its logarithm and converting the product to an integral. The echo shows a negligible dip close to the QCP for the 2-D DH; this again reflects the marginality of the situation to be discussed in the next section. On the other hand, one finds a sharp dip at the QCP in 1-D case (See Fig.~\ref{Fig:Echo_Massive_Dirac}).
The early time ($t\rightarrow0$) evolution of $\mathcal{L}(t)$ can be obtained analytically, and decays with time in a Gaussian manner, with a rate proportional to the phase space volume $(k_{max}L)^2$  near the QCP ($m \to 0$):
\begin{equation}
\mathcal{L}(t) = \exp{\left[-\frac{L^2 k_{max}^2 \delta^2 t^2}{4\pi}\right]}.
\label{Eq:Echo_Decay}
\end{equation}

\section{V. Generic behavior of the fidelity and the LE in arbitrary dimensions}
Both the ground state fidelity and the ground state LE of the 2-D DH do not display a sharp dip near the QCP, as opposed to the 1-D case. Instead, these quantities start dropping away from the QCP, when $m$ becomes comparable to $k_{max}$ (See Fig.~\ref{Fig:Fidelity_Massive_Dirac}). However, the fidelity susceptibility density still displays a noticeable peak at the QCP. We shall show below that this is a consequence of the marginality of the 2D case; for $d>2$, the fidelity and the LE will again show similar behavior, but the fidelity susceptibility will cease to show a peak at the QCP.

Both $\ln F$ and $\ln \mathcal{L}(t)$ can be written in the following form (see (\ref{Eq:Fidelity}) and (\ref{Eq:Echo_Massive_Dirac})):
\begin{equation}
G = \left(\frac{L}{2\pi}\right)^d\int_{-\sqrt{(2\pi/L)^2+m^2}}^{-\sqrt{k_{max}^2+m^2}} d\epsilon D(\epsilon) g(\epsilon),
\label{Eq:DOS}
\end{equation}
where $D(\epsilon)\propto|\epsilon|(\epsilon^2-m^2)^{d/2-1}$, is the density of states for the massive Dirac dispersion in $d$ dimensions, and $g(\epsilon)\leq0$, is the logarithm of the term involving the respective overlap of the states with energy $\epsilon$.

For $1/L\ll m \ll k_{max}$, one can work out how sensitive these quantities are to changes in $m$:
\begin{equation}
\frac{dG}{dm}\propto D(-m)g(-m)\propto\lim_{\epsilon\rightarrow -m}|\epsilon|(\epsilon^2-m^2)^{d/2-1}g(-m).
\end{equation}
For $d<2$, the density of states is singular at $\epsilon=-m$, implying a large sensitivity to small changes in $m$ near the QCP, and producing a sharp dip in the fidelity and LE. For $d=2$, $D(-m)=m$, drastically reducing the sensitivity of $G$ to changes in $m$ near the QCP and eliminating the sharp dip; however, $\chi_{F}\sim\ln F\sim G$ is still sensitive enough to $m$ to display a peak at the QCP (Figs.~\ref{Fig:Fidelity_Massive_Dirac} and \ref{Fig:Echo_Massive_Dirac}). For $d>2$, $D(-m)=dG/dm=0$, and there is neither a dip in the fidelity and LE nor a spike in $\chi_F$ as $m\rightarrow0$.

Very far away from the QCP ($m\gg k_{max}\gg1/L$), $G$ is trivially zero and both the fidelity and the LE are hence insensitive to $m$ and equal to unity. However, when $m\sim k_{max}\gg1/L$, $dG/dm\sim m^{d-1}g(-m)$ (a finite negative value) producing a drop in the quantities from unity. This drop grows stronger as the number of dimensions is increased.  It is to be noted that the nature of this drop is not
universal; rather it is a characteristic of the Dirac spectrum.

\section{VI. Conclusion}

In conclusion, the 2-D DH turns out to be an ideal model to explore the subtle effects of marginality on the ground state fidelity and the LE close to the QCP. The marginality is reflected in the absence of a sharp dip close to the QCP at $m=0$. The fidelity susceptibility shows a peak right at the critical point which becomes more prominent if the cutoff $k_{max}$ is increased (i.e., the high energy modes are incorporated). This already shows that unlike in 1-D, the high energy modes play a dominant role in the 2-D case, and hence the results crucially  depend on the cutoff $k_{max}$. Both the scaling of the fidelity susceptibility and the fidelity in the thermodynamic limit bear the signature of marginality: while the former shows a logarithmic scaling (with $L$ and $m$ in the appropriate limits)
there is a logarithmic correction in the latter. To explain the results from a general perspective, we consider a DP in an arbitrary number of dimensions. Analyzing the generic form of the logarithm of the fidelity (or the LE), and using the form of the density of states of the massive DH, we arrive at the following conclusions: for a DP in $d=2$ there is no sharp dip in the fidelity (in all limits) and the LE close to the QCP; the fidelity and the LE instead start to drop away from the QCP (when $|m| \sim k_{max}$); this behavior will persist for $d >2$. This drop (away from the critical point)  is negligibly small for $d=1$, where one observes a sharp dip at the QCP only. The fidelity susceptibility, on the other hand, shows a peak right at the QCP (thus enabling one to detect it) in the marginal case; however, this peak will be absent in dimensions higher than two. It would be interesting to probe similar effects of marginality for models with $\nu \neq 1$.

AAP acknowledges the KVPY fellowship and AD and SS acknowledge CSIR, New Delhi, for financial support.


\begin{thebibliography}{99}

\bi{sachdev99} S. Sachdev, {\it Quantum Phase Transitions} (Cambridge 
University Press, Cambridge, England, 1999). 

\bi{chakrabarti96} B. K. Chakrabarti, A. Dutta, and P. Sen, {\it Quantum Ising Phases and transitions in transverse Ising Models}, m41 (Springer, Heidelberg, 1996).

\bi{sondhi97} S. L. Sondhi, S. M. Girvin, J. P. Carini, and D. Shahar, Rev.  Mod. Phys. {\bf 69}, 315 (1997).


\bi{castroneto09} A. H. Castro Neto, F.  Guinea,  N. M. R. Peres, K. S. Novoselov and A. K. Geim, Rev. Mod. Phys. {\bf 81}, 109 (2009).

\bi{hasan10} M.H. Hasan, C.L. Kane, Rev. Mod. Phys. {\bf 82}, 3054 (2010).

\bi{qi11} Xiao-Liang Qi, Shou-Cheng Zhang, Rev. Mod. Phys. {\bf 83}, 1057 (2011).

\bi{novoselov2005two} K. Novoselov, A. Geim, S. Morozov, D. Jiang, M. Grig- orieva, S. Dubonos, and A. Firsov, Nature {\bf 438}, 197 (2005).

\bi{kane05} C. L. Kane and E. J. Mele, Phys. Rev. Lett. {\bf 95}, 226801 (2005).

\bi{bernevig06} B. Bernevig, T. Hughes, and S. Zhang, Science {\bf 314}, 1757 (2006).

\bi{quhe2012tunable} R. Quhe, J. Zheng, G. Lou, Q. Liu, R. Qin, J. Zhou, D. Yu, S. Nagase, W-N. Mei, Z. Gao, and J. Lu, NPG Asia Materials {\bf 4}, e6 (2012).

\bi{lee09} K. L. Lee, B. Gremaud,  R. Han, B-G Englert, and C. Miniatura, Phys. Rev. A {\bf 80}, 043411 (2009).

\bi{ochiai09}  T. Ochiai and M. Onoda,  Phys. Rev. B {\bf 80}, 155103 (2009).

\bibitem{zanardi06} P. Zanardi and N. Paunkovic, Phys. Rev. E {\bf 74}, 031123 (2006).

\bibitem{zhou08fidelity} H-Q. Zhou and J. P. Barjaktarevic, J. Phys. A: Math. Theor. {\bf 41} 412001 (2008). 

\bi{zhou08ground} H-Q. Zhou, R. Orus and Guifre Vidal, Phys. Rev. Lett. {\bf 100}, 080601 (2008).

\bi{sacramento11} P. D. Sacramento, N. Paunkovic and V. R. Vieira, Phys. Rev. A {\bf 84}, 062318 (2011).

\bi{grandi10} C. De Grandi, V. Gritsev and A. Polkovnikov, Phys. Rev. B {\bf 81}, 012303 (2010).

\bi{patel12} A. A. Patel and A. Dutta, Phys. Rev. B {\bf 86} 174306 (2012).

\bibitem{quan06} H. T. Quan, Z. Song, X. F. Liu, P. Zanardi, and C. P. Sun, Phys. Rev. Lett. {\bf{96}}, 140604 (2006).

\bi{damski12} B. Damski, arxiv: 1212.1825 (2012).

\bi{venuti07} L. Campos Venuti and P. Zanardi, Phys. Rev. Lett. {\bf{99}}, 095701 (2007), P. Zanardi, P. Giorda, and M. Cozzini, Phys. Rev. Lett. {\bf{99}}, 100603 (2007).

\bi{gu10} Shin-Jian Gu, Int. J. Mod. Phys. B {\bf 24}, 4371(2010).

\bi{dutta10} A. Dutta, U. Divakaran, D. Sen, B. K. Chakrabarti, T. F. Rosenbaum and G. Aeppli, arXiv:1012.0653 (2010).

\bi{gritsev09} V. Gritsev and A. Polkovnikov, arXiv:0910.3692 (2009), 
published in {\it Understanding Quantum Phase Transitions}, edited by 
L. D. Carr (Taylor and Francis, Boca Raton, 2010).

\bi{schwandt09} D. Schwandt, F. Alet, and S. Capponi, Phys. Rev. Lett. 
{\bf 103}, 170501 (2009); A. F. Albuquerque, Fabien Alet, Clement Sire, and Sylvain Capponi, Phys. Rev. B {\bf 81}, 064418 (2010).\bi{mukherjee11} V. Mukherjee,  A. Polkovnikov and A. Dutta,  Phys. Rev. B {\bf 83}, 075118 (2011)

\bi{mukherjee11a} V. Mukherjee, A. Dutta and D. Sen, Phys. Rev. B {\bf 85}, 024301 (2012).

\bi{thakurati11} M. Thakurathi, D. Sen and A. Dutta, Phys. Rev. B {\bf 86}, 245424 (2012).

\bi{rams11} M. M. Rams and B. Damski, Phys. Rev. Lett. {\bf 106}, 055701 
(2011); M. M. Rams and B. Damski, Phys. Rev. A {\bf 84} 032324 (2011).

\bi{echo}F. M. Cucchietti, S. Fernandez-Vidal, and J. P. Paz, Phys. Rev. A {\bf 75}, 032337 (2007); D. Rossini, T. Calarco, V. Giovannetti, S. Montangero, and R. Fazio, ibid. {\bf 75}, 032333 (2007); L. Campos Venuti and P. Zanardi, Phys. Rev. A {\bf 81}, 022113 (2010);
L. Campos Venuti, N. T. Jacobson, S. Santra, and P. Zanardi, Phys.
Rev. Lett. {\bf 107}, 010403 (2011).

\bi{damski11} B. Damski, H. T. Quan, and W. H. Zurek, Phys. Rev. A 83, 062104
(2011); T. Nag, Uma Divakaran and A. Dutta, Phys. Rev. B {\bf 86}, 020401 (2012); V. Mukherjee, S. Sharma, and A. Dutta
Phys. Rev. B {\bf 86}, 020301 (2012). 

\bi{zhang09} J. Zhang, F. M. Cucchietti, C. M. Chandrashekar, M. Laforest, C. A. Ryan,
M. Ditty, A. Hubbard, J. K. Gamble, R. Laflamme, Phys. Rev. A {\bf 79}, 012305
(2009).

\end{thebibliography}
\end{document}